# Effectiveness of SCADA System Security Used Within Critical Infrastructure

Joshua Taylor
*Bournemouth University*
Bournemouth, United Kingdom
s5085242@bournemouth.ac.uk
https://orcid.org/0000-0003-1277-855X

*Abstract*—Since the 1960s Supervisory Control and Data Acquisition (SCADA) systems have been used within industry. Referred to as critical infrastructure (CI), key installations such as power stations, water treatment and energy grids are controlled using SCADA. Existing literature reveals inherent security risks to CI and suggests this stems from the rise of interconnected networks, leading to the hypothesis that the rise of interconnectivity between corporate networks and SCADA system networks pose security risks to CI. The results from studies into previous global attacks involving SCADA and CI, with focus on two highly serious incidents in Iran and Ukraine, reveal that although interconnectivity is a major factor, isolated CIs are still highly vulnerable to attack due to risks within the SCADA controllers and protocols.

*Keywords—SCADA, networks, Critical, Infrastructure, Interconnectivity, Security, Safety,*

## I. INTRODUCTION

Supervisory Control and Data Acquisition (SCADA) systems are commonly used to monitor and control Critical Infrastructure (CI) such as airports, oil fields and power plants [1]. SCADA system networks were designed before the interconnected design of modern system networks. CI systems are crucial to economic and physical safety and security but increasing demand for interconnectivity over recent decades within CI industries presents an array of security risks to SCADA system networks. By studying the impacts of recent high-profile attacks against critical infrastructure, this work intends to explore the security implications of the SCADA control networks. The aim is to form a conclusion to the hypothesis presented from studying related literature; the continuing rise of interconnectivity between corporate networks and SCADA system networks pose security risks to CI.

## II. SCADA WITHIN CRITICAL INFRASTRUCTURE

### A. SCADA Network Systems

SCADA refers to the technology enabling a user to control instructions and collect data between multiple distant facilities, the name SCADA is derived from the acronym meaning "supervisory control and data acquisition" [2]. SCADA systems have been in use ever since industrial control systems were first utilised [3]. Although there is no definitive historical moment where such systems were first used, existing literature indicates they begun in the 1960's [4],[5].

The primary benefit of SCADA is to reduce efforts and subsequent costs of monitoring dispersed facilities that require frequent and immediate intervention but present extreme logistical effort to visit regularly [2]. Typically, SCADA systems are centred around a system controller known as the Master Terminal Unit (MTU) used via Human Computer Interface (HCI) by an 'operator' [2]. In a SCADA system the MTU communicates data between one or more otherwise located Remote Terminal Units (RTU) via signal modulator and demodulator (MODEM) across a physical landline [2]..

### B. Significance of Critical Infrastructure

SCADA systems serve a multitude of uses in geographically widespread industry including but not limited to oil fields, pipelines and powerplants [2],[3],[5]. Furthermore, car park monitoring and building control systems used in running and maintaining airports are operated via SCADA [6]. Such facilities, including their systems and networks, are crucial to the safety, security and economic interests of people and organisations so are considered CIs [1]. By this definition airports are considered as CI facilities, so without properly functioning infrastructure control there arises significant operational and public safety concerns [7].

The crucial nature of CIs along with their large attack surface presents them as prime targets for state actors and crime groups [8]. Most CIs are monitored and controlled with SCADA systems resulting in 'devastating consequences' in event of a cyber-attack against said SCADA systems [1]. During the very early development of SCADA protocols, network security was of no concern and design requirements instead emphasised higher performance on meeting task constraints on the network [9]. But in the last few decades there has been growing demand to increase interconnectivity between facilities and corporate networks [9], resulting in more complex software and remote configuration [10]. This increases the number of possibilities for exploitation and intrusion [10]..

## III. RELATED WORK

Study of related work in this area is broken into three trend categories: the implications of interconnectivity, the vulnerabilities in the design of SCADA systems and the previously proposed security measures to mitigate these risks.

### A. Adapting to Interconnectivity within Modern Industry

An influential 2007 conference paper stated that standard internet communication protocols (TCP/IP) had become prominent in modern SCADA networks and argues that its facilitation of interconnectivity raises serious issues [11]. Supported later by Pliatsios *et al.* [1] which confirms the rise of interconnectivity in industrial applications resulted in increased connection between SCADA systems, corporate networks and the internet but also introduced greater security risks, however this survey serves only as a report on existing trends. Nevertheless, the findings of a 2017 conference paper which agrees interconnectivity leads to increased interest from attackers, determined that operators at the time didn't physically separate their industrial networks and thus still fail to provide basic security [10]. Despite this, interconnectivity has some advantages including remote accessibility for maintenance, the flow of constant information [10] and subsequently reduces operation costs [1].

### B. Inherent Vulnerability of SCADA Systems and Networks

SCADA systems evolved from hardware and software platforms used in the 1960s [4] thus originally employed primitive serial protocols and communications infrastructures [11]. The self-contained nature of SCADA systems suggested they were externally secure at the time [4] and so focus was placed on operational requirements rather than security [11]. Even since the emergence of the internet, SCADA systems were assumed to be less vulnerable to IT-system attacks [4], this resulted in protocols being designed without security considerations [11]. Igure *et al.* [9], an older yet still highly relevant article, argues that SCADA protocols typically don't support cryptographic techniques, this is supported by a recent survey stating both Modbus and DNP3 protocols used for SCADA networks do not support access mechanisms and use unencrypted communication [1]. Furthermore, Igure *et al.* [9] claims the use of real-time operating systems (RTOS) contributes to security issues since demonstrations showed RTOS being vulnerable to denial of service (DoS) attacks, suggesting SCADA networks may be open to such attacks. This argument is strengthened by a later article which presents DoS and Unauthorised access as being the main threats to SCADA [12]. Finally, Igure *et al.* [9] introduces the idea that the use of Commercial Off-The-Shelf (COTS) software and hardware contributes to SCADA vulnerability, claiming COTS software is often insecure.

### C. Proposed SCADA Security Measures and Practises

A popular and influential book by Krutz [4] titled 'Securing SCADA Systems' outlines how SCADA systems can be protected by applying modified IT-system security methods, SCADA system security policies, vulnerability analysis, risk management, personnel awareness and elimination of unnecessary network connections. A journal article by Nicholson *et al.* [12] states specific systems (firewalls, antivirus etc), mock attacks, honeynets (simulated SCADA systems to monitor hacker trends) and penetration testing are among the currently employed preventative measures. It concludes that their findings indicate the need for further work in the area of SCADA security [12], Pliatsios *et al.* [1] supports this conclusion and adds that malware is constantly being developed by adversaries, it goes on to state that using cryptographic mechanisms should be enforced along with deploying intrusion detection systems. The survey [1] references Chandia *et al.* [11] whose article proposes two strategies which it claims have minimal impact on real-time industrial operations. The first involves the deployment of a security services suite that provide risk mitigation facilities in response to identified risks and known vulnerabilities, the second proposes a forensic system for the capture and analysis of SCADA network traffic [11].

### D. Discussion

Study of related work outlines the security issues caused by the modern industrial adoption of interconnectivity standards, explores existing vulnerabilities and proposes security measures in industrial SCADA networks. Investigation into the adoption of interconnectivity, despite highlighting its advantages, concluded that security risks in modern SCADA systems and networks are as a direct result. The vulnerabilities in SCADA networks are primarily due to their originally isolated nature, as a result security was not considered in the design of SCADA protocols yet increases in development towards interconnectivity have since made it necessary. Multiple sources propose and present an array of countermeasures and practises to better mitigate the risks and impacts of these security issues however the literature conclusively agrees that further work needs to be conducted in this area to better reduce threats to CI. The related work determines that interconnectivity exposes the inherent vulnerabilities of SCADA systems, leading to the hypothesis of this study: that the continuing rise of interconnectivity between corporate networks and SCADA system networks pose security risks to critical infrastructure.

## IV. AIM OF THE STUDY

The aim of this work is to test the presented hypothesis. Leading on from the related work section, the research is broken into three objectives: To investigate into the exploitation methods of recent high-profile attacks against critical infrastructure and the effectiveness of the implemented security measures; To add context, study the impacts of these attacks to determine the extent of the risks associated with interconnectivity; Research the lessons learned from the attack case studies that determined further security measures to prevent similar attacks in the future. The research will conclude as to whether interconnectivity is the primary cause of these incidents and form an answer to the hypothesis.

## V. STUXNET ATTACK AGAINST IRANIAN NUCLEAR FACILITY

In July 2010 the Natanz uranium enrichment plant in Iran fell victim to Stuxnet [13], a worm malware that experts perceive as being the first case of a cyber weapon [14]. Over 60,000 computers were infected across the world, most of them in Iran [13]. Stuxnet was intended to physically damage or destroy a military target and was aimed at industrial controllers, some of which were part of a SCADA system [15].

Using off-the-shelf resources from the global cyber-crime community [13], it was targeted specifically at controllers from the Siemens manufacture by checking the model numbers, configurations and code [15]. When a PC infected with Stuxnet connected to a Siemens controller the malware used Siemens default passwords to access the windows operating system [13]. It then loaded code into the controller [15] by replacing the controllers original .dill file with its malicious one [14]. Designed as a stealth feature, Stuxnet remained dormant as the controller continued to function normally, occasionally taking over to call malicious commands [15]. In the case of Natanz, although denied officially by Iran [14], those commands adjusted the electrical current of the uranium centrifuges, causing them to alternate their speed contrary to design and damaging them

[13]. Damage was not exclusive to Iran; an Indian satellite was also infected with Stuxnet [13].

It's not clear exactly how the worm spread [13] but its creators anticipated that the target wouldn't be accessible via the internet [14]. Unlike a conventional worm, Stuxnet relied on local distribution methods such as local networks and USB sticks [14], [15], although Stuxnet was also able to spread via the internet [13]. The primary vulnerability was that controllers don't verify the integrity of their programming code, to resolve this, system owners must replace the installed systems [15]. Iran was able to utilise the global cyber community to source solutions to the worm [13] but this attack shows that isolation from the internet does not guarantee a secure SCADA system [14], which concludes that interconnectivity was not a contributing factor to Stuxnet.

VI. BLACKENERGY ATTACK ON THE UKRAINIAN POWER GRID

In December 2015, a foreign attacker remotely disrupted a SCADA system [16], this resulted in power outages in three provinces across Ukraine which affected over 225,000 homes for almost six hours [17]. The attack wasn't specific to Ukrainian infrastructure and the methods it used could apply to other global infrastructures with varying degrees of impact [16]. Initially spread using email spear phishing which tricked corporate workstation users into opening the attached document that installed Black Energy 3 (BE3) [17]. BE3 provides the attackers with a HTTP based bot with control functionalities [18] and allowed them to discover the active directory, leading to a brute-force attack on the stored passwords [17]. The stolen passwords were then used to access the SCADA workstations and servers to control the circuit breakers [16].

The incident affected more than 50 substations and impacted approximately 225,000 customers, requiring technicians to be sent to the substations to control the power system manually [17]. The attackers used malware known as KillDisk to wipe the master boot records of the infected systems, ensuring that the substations could not be remotely restored [16]. Interconnectivity was a factor in this attack, if the SCADA system is connected to enterprise networks then it can be infected by malware like BE3 [17]. Although BE3 initially infiltrated the enterprise systems via social engineering, the firewall rules were improperly configured on the router that separates the networks [17], thus the attackers used virtual private networks (VPN's) to remotely connect to the SCADA network and issue commands [16].

Unlike Stuxnet, this attack was against solely civilian infrastructure [16]. Several critical infrastructures including nuclear plants and electric grids, had been compromised by BlackEnergy since 2011 [18]. Mitigation for this focuses on limiting remote access, segmenting networks with properly configured firewalls and limiting remote connections to only required personnel [16]. The attack surface would have been reduced by limiting control to required workstations and insuring control systems are not connected to the internet [17].

VII. IMPACT OF ATTACKS ON SCADA SYSTEMS

There is potential for harm to life and the economy if SCADA systems are breached [4]; In 2000 a disgruntled former Queensland water treatment employee gained unauthorised access into the management system, causing raw sewage to spill out into local parks and rivers [1][6][12]. This and the other case examples support modern claims that security considerations for SCADA systems are gaining higher priority within industry due to the potential impact on the physical safety of employees, customers and communities [1]. Since 2001 almost 70% of incidents were from attacks outside the SCADA network [9].

A 2017 paper analysed different types of SCADA attack with case examples and concludes that attacks on industrial infrastructure primarily aim at sabotaging production using highly sophisticated malware [10]. Although supported with examples, this conclusion is slightly weakened by prior suggestions that many businesses are reluctant to publicize attacks for fear of disrepute in exposing any potential weaknesses, resulting in difficulty to determine whether risks were due to general IT vulnerabilities or due to SCADA system related vulnerabilities [12].

VIII. CONCLUSION

Existing literature concluded that there were serious threats surrounding SCADA controlled CI and conclusively agreed that further work needed to be done in this area. Studying previous attacks on CIs and their impact further highlights the seriousness of these risks and the significance of exploiting SCADA Systems.

The use of interconnected networks within Ukrainian enterprise infrastructures along with inadequate firewall segmentation led to a disruptive power outage across a section of the country. The case study into Stuxnet revealed that disruption and damage of SCADA controlled systems is still a possibility even within isolated networks, but it requires in-depth knowledge and access to resources. Although interconnectivity was a largely contributing factor to the attack against the Ukrainian power grid, the hypothesis loses strength in the case of the Stuxnet attack against Iran.

Overall, the hypothesis is valid due to consistently ongoing security implications stemming from the interconnected nature of modern industrial operations. However, it is important to understand that resolving interconnectivity issues does not guarantee secure CI, there are security risks within the SCADA systems themselves that need resolution and complacency regarding isolated systems is ill advised.